\documentclass[smallextended]{svjour3}     
\smartqed  
\usepackage{graphicx}
 \journalname{Gen Relativ Gravit}
\begin{document}

\title{Gravitational interaction of antimatter
}


\author{Massimo Villata
}


\institute{M.\ Villata \at
              INAF, Osservatorio Astronomico di Torino, Italy\\
              \email{villata@oato.inaf.it}           
}

\date{Submitted: 2 February 2010}

\maketitle

\begin{abstract}
Until now, there is no experimental evidence on the gravitational behaviour of antimatter. While we may be confident that antimatter attracts antimatter, we do not know anything on the interaction between matter and antimatter. We investigate this issue on theoretical grounds. Starting from the CPT invariance of physical laws, we transform matter into antimatter in the equations of both electrodynamics and gravitation. In the former case, the result is the well-known change of sign of the electric charge. In the latter, we find that the gravitational interaction between matter and antimatter is a mutual repulsion. This result supports cosmological models attempting to explain the Universe accelerated expansion in terms of a matter-antimatter symmetry.
\keywords{gravitation \and antimatter \and cosmology: theory \and dark energy \and large-scale structure of Universe}
\end{abstract}

\section{Introduction}
\label{intro}
The discovery of antimatter (in 1932) raised the question about its reaction to a gravitational field. Until now no clear experimental answer could be obtained, due to the weakness of gravitation compared to the electromagnetic forces governing antiparticle motion in accelerators, and, even when dealing with electrically neutral antihydrogen, due to its fast annihilation with matter. 

While CPT invariance of physical laws assures that antimatter is gravitationally attracted by antimatter exactly as matter by matter, no definitely convincing theoretical argument has been so far proposed to discriminate whether matter and antimatter mutually attract or repel. 

Here we show that the answer can be found in the equations of general relativity, when the above mentioned CPT operation is properly applied. The result is that matter and antimatter repulse each other. 

Besides the obvious importance of this result in terms of pure knowledge, it also supports the recent attempts to explain the observed accelerated expansion of the Universe with matter-antimatter models, which look much simpler than the $\Lambda$-CDM model, that is based on the dominant existence of the so-called `dark energy', of unknown origin.

\section{Method and results}
\label{sec:1}
Physical laws are known to be invariant under the combined CPT operations, where C (charge conjugation) is the particle-antiparticle interchange, P (parity) is the inversion of the spatial coordinates, and T is the reversal of time. Dealing with classical particles and antiparticles, it is
\begin{equation}
{\rm CPT\,:}\qquad{\rm d}x^\mu\,\rightarrow\,-{\rm d}x^\mu\,,\qquad q\,\rightarrow\,-q\,,
\end{equation}
where $q$ is the electric charge.

According to the Feynman-St\"uckelberg interpretation \cite{Ref1,Ref2,Ref3}, antiparticles are nothing else than the corresponding particles travelling backwards in time. Hence, when we deal with a physical system composed of both matter and antimatter, the correct transformation to be applied to the time-reversed component for treating it in our time direction together with normal matter is the CPT one. Even if one would not believe in the Feynman-St\"uckelberg interpretation, this is in any case the only law-invariant transformation for replacing matter with antimatter in a given physical system.

We first see how this works in electrodynamics, then we apply to gravitation. Throughout this paper, we use standard notations and symbols, units with $c=1$, and a $(-,+,+,+)$ metric. 

\subsection{Electrodynamics}
\label{sec:2}
The equation of motion, i.e.\ the Lorentz force law, describes the dynamics of a particle of charge $q$ in an external electromagnetic field $F^{\mu\nu}$, and it is (obviously) invariant under CPT:
\begin{equation}
{\rm CPT\,:}\qquad{{\rm d}^2x^\mu\over{\rm d}\tau^2}={q\over m}g_{\nu\lambda}F^{\mu\lambda}{{\rm d}x^\nu\over{\rm d}\tau}\quad\rightarrow\quad{-{\rm d}^2x^\mu\over{\rm d}\tau^2}={-q\over m}g_{\nu\lambda}(-F^{\mu\lambda}){-{\rm d}x^\nu\over{\rm d}\tau}\,,
\end{equation}
which means that the corresponding time-reversed, i.e.\ antimatter, system obeys the same law as the original one, with charge and field reversed in sign. For example, an electron in a proton field (hydrogen) behaves exactly as a positron in an antiproton field (antihydrogen).

If we apply CPT to the particle only (and not to the field), we have the equation for the corresponding antiparticle interacting with the original (non-time-reversed) field:
\begin{equation}
{{\rm d}^2x^\mu\over{\rm d}\tau^2}=-{q\over m}g_{\nu\lambda}F^{\mu\lambda}{{\rm d}x^\nu\over{\rm d}\tau}\,,
\end{equation}
i.e.\ the equation for a particle of opposite charge, as expected (e.g., following the example above, a positron repulsed by a proton field). If we now CPT-transform all this last equation (or, which is the same, CPT-transform only the field in Eq.\ (2)), we get the same equation, but now governing the motion of a normal particle in a time-reversed field (generated by antiparticles), namely, now the minus sign comes from the field and not from the charge.

We can further check what happens with the inhomogeneous Maxwell equations for the electromagnetic field produced by a given four-current density:
\begin{eqnarray}
{\rm CPT\,:}\qquad{\partial\sqrt{g}F^{\mu\nu}\over\partial x^\mu}=-\sum_n{q_n\int{\delta^4(x-x_n){{\rm d}x_n^\nu\over{\rm d}\tau_n}{\rm d}\tau_n}}\quad\nonumber\\
\rightarrow\quad{\partial\sqrt{g}(-F^{\mu\nu})\over -\partial x^\mu}=-\sum_n{-q_n\int{\delta^4(x-x_n){-{\rm d}x_n^\nu\over{\rm d}\tau_n}{\rm d}\tau_n}}\,.
\end{eqnarray}
Although the charges and the field are changed in sign with respect to their matter counterparts, the equation is invariant, as expected. In particular, the minus signs from the charges and the four-velocities cancel, so that the current density on the right-hand side is CPT-even. The field is CPT-odd, and the electromagnetic potential $A_\mu$, related to the field by
\begin{equation}
F_{\mu\nu}={\partial A_\nu\over\partial x^\mu}-{\partial A_\mu\over\partial x^\nu}\,,
\end{equation}
is CPT-even (it would be the opposite under PT alone).

Thus, in electrodynamics this procedure works, yielding the expected interactions for particles and antiparticles.

\subsection{Gravitation}
\label{sec:3}
In the gravitational equations of the general theory of relativity, one of the most evident mathematical differences is that all tensor ranks of potentials ($g_{\mu\nu}$), fields ($\Gamma^{\lambda}_{\mu\nu}$), and currents ($T^{\mu\nu}$) are increased by one. This matches the fact that here the `charge' is no longer a scalar, but the energy-momentum four-vector $p^\mu=m\,{\rm d}x^\mu/{\rm d}\tau$. Thus, when applying CPT to gravitation, C is ineffective, but the `charge' changes sign by PT. Indeed, by (C)PT-transforming the energy-momentum current density $T^{\mu\nu}$,
\begin{equation}
{1\over\sqrt{g}}\sum_n{m_n\int{\delta^4(x-x_n){{\rm d}x_n^\mu\over{\rm d}\tau_n}{{\rm d}x_n^\nu\over{\rm d}\tau_n}{\rm d}\tau_n}}\quad\rightarrow\quad{1\over\sqrt{g}}\sum_n{m_n\int{\delta^4(x-x_n){-{\rm d}x_n^\mu\over{\rm d}\tau_n}{-{\rm d}x_n^\nu\over{\rm d}\tau_n}{\rm d}\tau_n}}\,,
\end{equation}
we see that it is (C)PT-even, as expected from its rank, due to the change in sign of both the four-velocity and the `charge', similarly to the electromagnetic current density in Eq.\ (4). Moreover, potentials ($g_{\mu\nu}$) and fields ($\Gamma^{\lambda}_{\mu\nu}$) are (C)PT-even and odd, respectively (according to their ranks, even if the latter is not a tensor), as in electrodynamics.

In the Einstein field equation
\begin{equation}
R_{\mu\nu}-{1\over 2}g_{\mu\nu}R=-8\pi GT_{\mu\nu}\,,
\end{equation}
both sides are clearly even, again as in Eq.\ (4). This (C)PT invariance of the field equation implies that an antimatter energy-momentum tensor generates a gravitational field (or space-time curvature) in the same way as matter does (but with inverted `charges' and fields).

As a consequence of the equivalence principle, in the equation of motion (i.e.\ the geodesic equation) the mass disappears. However, in the following it may be useful to keep the ratio $m_{({\rm g})}/m_{({\rm i})}=1$ visible in the equation:
\begin{equation}
{{\rm d}^2x^\lambda\over{\rm d}\tau^2}=-{m_{({\rm g})}\over m_{({\rm i})}}{{\rm d}x^\mu\over{\rm d}\tau}\Gamma^{\lambda}_{\mu\nu}{{\rm d}x^\nu\over{\rm d}\tau}\,.
\end{equation}
As in Eq.\ (2), the acceleration is odd, the `charge' $p^\mu=m_{({\rm g})}{\rm d}x^\mu/{\rm d}\tau$ is odd, as well as the field and the four-velocity, and the equation is (C)PT invariant. Therefore, an anti-apple would fall onto the head of an anti-Newton walking on an anti-Earth, exactly in the same way as it happened here some time ago.

What about an anti-apple on the Earth, or an apple on an anti-planet? As in the electrodynamic case of Eq.\ (2), we must (C)PT-transform one of the two components, no matter which one.
The result is a gravitational acceleration with the opposite sign, i.e.\ a repulsion between matter and antimatter:
\begin{equation}
{{\rm d}^2x^\lambda\over{\rm d}\tau^2}=-{-m_{({\rm g})}\over m_{({\rm i})}}{{\rm d}x^\mu\over{\rm d}\tau}\Gamma^{\lambda}_{\mu\nu}{{\rm d}x^\nu\over{\rm d}\tau}\,,
\end{equation}
where the change of sign comes from either the (C)PT-odd field (anti-Earth) or the (C)PT-odd `charge' (anti-apple). Comparing with Eq.\ (8), the gravitational repulsion may be seen as the result of an effective negative gravitational mass.

\section{Conclusions}
\label{sec:4}
This theoretical derivation of the gravitational repulsion between matter and antimatter supports cosmological models attempting to explain the observed accelerated expansion of the Universe through such a repulsion between equal amounts of the two components. Very promising appears to be the model by Benoit-L\'evy \& Chardin \cite{Ref4,Ref5}.

The gravitational repulsion would prevent the mutual annihilation of isolated and alternated systems of matter and antimatter. The location of antimatter could be identified with the well-known large-scale (tens of Mpc) voids observed in the distribution of galaxy clusters and superclusters. Indeed, Piran \cite{Ref6} showed that these voids can originate from small negative fluctuations in the primordial density field, which ({\it acting as if they have an effective negative gravitational mass}) repel surrounding matter, and grow as the largest structures in the Universe. These new cosmological scenarios could eliminate the uncomfortable presence of an unidentified dark energy, and maybe also of cosmological dark matter, which, according to the $\Lambda$-CDM concordance model, would together represent more than the 95\% of the Universe content. 

If large-scale voids are the location of antimatter, why should we not observe anything there? There is more than one possible answer, which will be investigated elsewhere.

\begin{acknowledgements}
The author wishes to thank Gabriel Chardin, Marco Raiteri, and Claudia M.\ Raiteri for stimulating discussions.
\end{acknowledgements}


\begin{thebibliography}{}
%
%
\bibitem{Ref1}
St\"uckelberg, E.C.G.: La m\'ecanique du point mat\'eriel en th\'eorie del relativit\'e et en th\'eorie des quanta. Helv.\ Phys.\ Acta 15, 23--37 (1942)
\bibitem{Ref2}
Feynman, R.P.: A relativistic cut-off for classical electrodynamics. Phys.\ Rev.\ 74, 939--946 (1948)
\bibitem{Ref3}
Feynman, R.P.: The theory of positrons. Phys.\ Rev.\ 76, 749--759 (1949)
\bibitem{Ref4}
Benoit-L\'evy, A., Chardin, G.: Observational constraints of a symmetric Milne universe. In: Proceedings of the 43rd Rencontres de Moriond 2008, arXiv:0811.2149v1 [astro-ph] (2008)
\bibitem{Ref5}
Benoit-L\'evy, A., Chardin, G.: Do we live in a ``Dirac-Milne'' Universe? arXiv:0903.2446v1 [astro-ph.CO] (2009)
\bibitem{Ref6}
Piran, T.: On gravitational repulsion. Gen.\ Relativ.\ Gravit.\ 29, 1363--1370 (1997)
\end{thebibliography}


\end{document}